\documentclass[fleqn,10pt]{wlscirep}
\usepackage[utf8]{inputenc}
\usepackage[T1]{fontenc}

\title{Adaptive Radiofrequency Shimming in MRI using Reconfigurable Dielectric Materials}

\author[1]{Paulina \v{S}iuryt\.{e}}
\author[1,+]{Robert van de Velde}
\author[1,+]{Jasper van Leeuwen}
\author[2]{\"{O}mer Can Akg\"{u}n}
\author[3]{Wyger Brink}
\author[1,*]{Sebastian~Weing\"{a}rtner}
\affil[1]{Department of Imaging Physics, Delft University of Technology, Delft, 2628 CJ, the Netherlands}
\affil[2]{Nikhef, Amsterdam, 1098 XG, the Netherlands}
\affil[3]{Department of Magnetic Detection and Imaging, TechMed Centre, University of Twente, Enschede, 7522 NB, the Netherlands}

\affil[*]{corresponding author (s.weingartner@tudelft.nl)}

\affil[+]{these authors contributed equally to this work}


\begin{abstract} 
Inhomogeneity of the transmitted radiofrequency field ($B_1^+$) is a major factor hindering the image quality in Magnetic Resonance Imaging (MRI) at high field strengths. 
Here, a novel approach is presented, to locally modulate the $B_1^+$ utilizing an array of high permittivity materials with switchable connections. 
A 3$\times$3 array of barium titanate suspension elements was constructed, with two PIN diode-based switchable connectors per element. 
Electromagnetic simulations were performed to determine configurations that produce strong $B_1^+$ modulation. 
Remote $B_1^+$ field switching was tested in a disk- and and a torso-shaped phantom at 3T by applying different bias voltages to the PIN diodes. 
The attained $B_1^+$ modulation was assessed at various switching pattern positions and various depths within the phantoms.
The configuration with the strongest effect size has produced up to 11$\%$ modulation in simulations at 15 mm depth, with excellent translation properties. 
The effects were successfully replicated in phantoms, with a 5 V bias voltage producing up to 11.6$\pm$0.2$\%$ modulation. 
At the relative depth of the human heart, up to 6$\%$ of modulation was observed in the torso phantom.
The presented method may provide a promising direction for cost-effective, and adaptive $B_1^+$ shimming without changes to the scanner hardware.
\end{abstract}
\begin{document}

\flushbottom
\maketitle

\thispagestyle{empty}

\section*{Introduction}

In Magnetic Resonance Imaging (MRI), high static magnetic field strengths ($B_0$) of 3T and above have been increasingly used due to the gains in available signal-to-noise ratio (SNR) \cite{bernstein2006}. However, image quality at high-field MRI can suffer from artifacts caused by inhomogeneities in the transmitted radiofrequency (RF) field, $B_1^+$ \cite{kataoka2007}. These artifacts become apparent when the RF wavelength approaches the dimensions of the anatomy of interest, such as in body imaging at 3T or neuroimaging at 7T and above \cite{kataoka2007,soher2007}. In these configurations, the formation of standing waves decreases the transmit field homogeneity, giving rise to image shading and contrast non-uniformities. In cardiac MRI, this can lead to signal variations of up to 50$\%$ across the heart \cite{sung2008}, reducing image quality and hindering clinical interpretation \cite{gutberlet2004}. 

Several methods have emerged to improve $B_1^+$ field homogeneity, a process referred to as shimming. Parallel RF transmission (pTx) can be used when multiple RF transmit channels are available \cite{deniz2019}. The amplitude and phase of each channel can be configured to improve the homogeneity of the resulting superimposed $B_1^+$ field. This has become the technological standard in many body imaging applications at 3T \cite{brink2015a}. Additionally, individual control of the transmitted waveform can enable short-time scale variation of RF pulses to improve magnetization homogeneity further \cite{wald2009,williams2023}. However, in both cases, the shimming capabilities are limited by the available software and hardware \cite{vorobyev2020}. As dual-channel transmit systems are offered by most vendors as part of top-of-the-line system architecture and require expensive upgrades or replacements in conventional implementations, pTx does not always provide a sustainable solution for existing and older systems.

An alternative for $B_1^+$ shimming is the use of dielectric materials, often termed "dielectric pads" or "dielectric shimming". These pads typically consist of materials with high relative permittivity ($\epsilon_r$) such as water, titanate suspensions \cite{teeuwisse2012, webb2011} or solid ceramics \cite{koolstra2017, webb2022}. These pads are placed on the subject close to the area of interest. During RF transmission a displacement current is induced in the high permittivity material, which in turn, gives rise to a secondary magnetic field. This field is superimposed onto the field generated by the RF transmit coil, improving the $B_1^+$ field homogeneity in the artifact-affected areas \cite{brink2015}. Several studies have demonstrated promising improvements of the $B_1^+$ homogeneity using barium titanate suspension pads in body MRI at 3T \cite{brink2014,heer2016}. Although dielectric pads offer a cheap solution to improve $B_1^+$ homogeneity, the optimal design of the pad geometry and dielectric properties is application- and, ultimately, subject-specific. As a result, the clinical use of the dielectric pads is limited to the use of a single "one-size-fits-all" design, involving compromises on the RF shimming performance.

In this work, we aim to improve the adaptability of dielectric shimming using an array of small dielectric elements that are electrically interconnected. The connections between the elements can be switched using PIN diodes to change the $B_1^+$ field modulation. Electromagnetic (EM) simulations were performed to optimize the configuration and to study the effect of varying the dielectric permittivity. A 3D-printed array of cavities filled with a high permittivity titanate suspension was used for in-situ $B_1^+$ shimming experiments at 3T. The $B_1^+$ field modulation was measured in two different phantoms, using various connection patterns and multiple depths below the phantom surface.

\section*{Methods}

\begin{figure}[ht]
    \centerline{\includegraphics[width=0.6\textwidth]{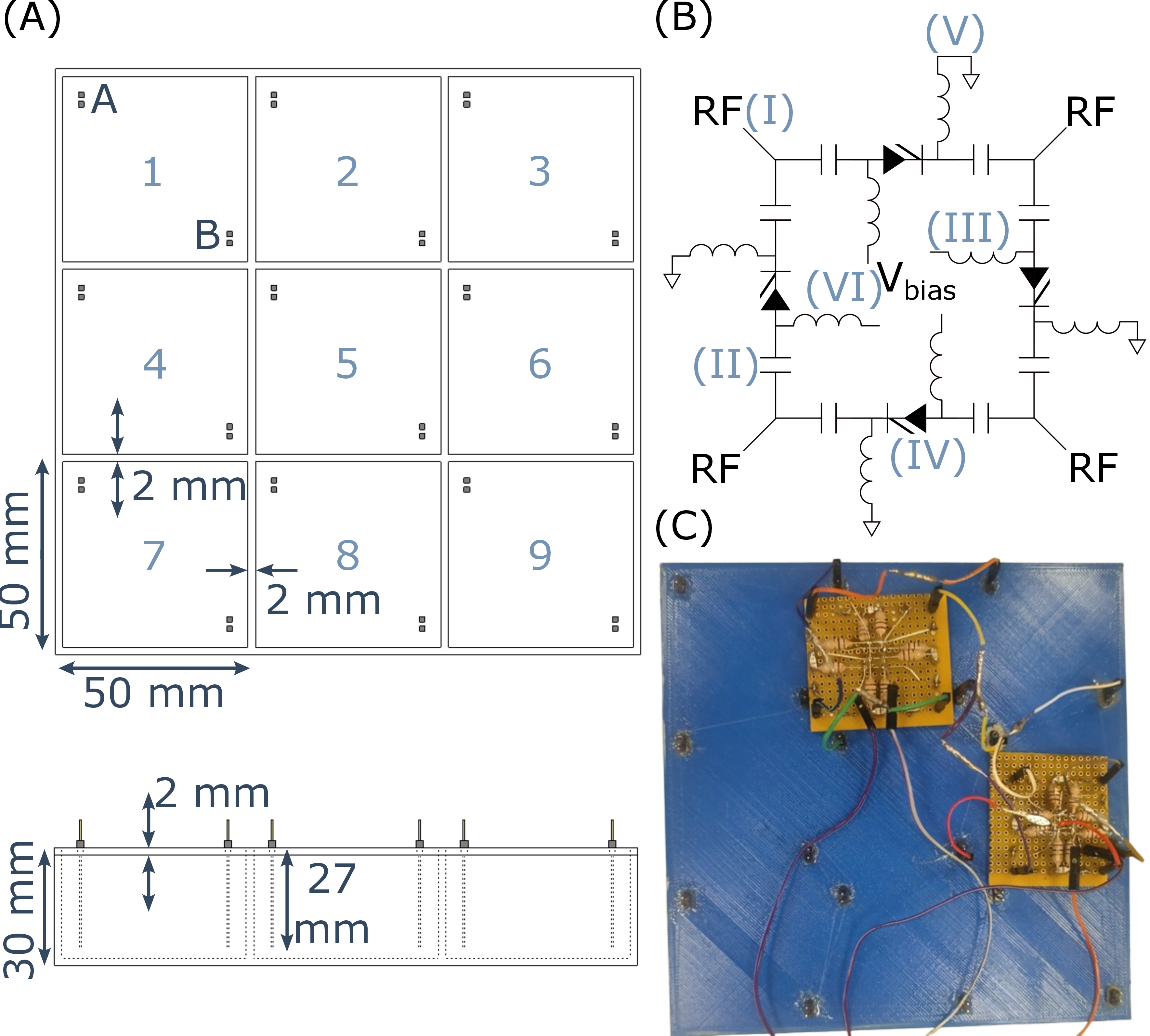}} 
    \caption{Switchable dielectric shimming prototype. (A) Schematic overview of the 3$\times$3 array of dielectric elements with a single element width of 50 mm, height 30 mm, electrode length 27 mm, and inter-element gap width of 2 mm. The array elements and connectors are labeled (1-9, A-B) in blue for reference purposes. (B) Schematic diagram of the RF switching circuit for interconnecting four elements of the dielectric array. (I) Header pin connector. (II) DC choke. (III) RF choke. (IV) PIN diode. (V) Ground. (VI) Bias voltage ($V_{bias}$). (C) 3D-printed polylactic acid container, filled with a 25$\%$ v/v barium titanate slurry (estimated relative permittivity of 165). An epoxy layer was applied on the inside for waterproofing, and the container was sealed using universal adhesive. The switching mechanism is shown connecting four adjacent elements in parallel (2A-3A-6A-5A-2A and 2B-3B-6B-5B-2B). The bias voltage is applied to each printed circuit board using a pair of jumper wires (visible at the bottom).}
    \label{fig1}
\end{figure}

The proposed approach is based on multiple small, sub-wavelength-sized dielectric elements, which can be electrically coupled using switchable connections to modulate the $B_1^+$ field. To this end, a prototype device was designed as a 3$\times$3 array of cubical cavities (50 mm$\times$50 mm$\times$30 mm$^3$, 2 mm separation, Fig.~\ref{fig1}A). Each cavity is electrically insulated and filled with a high permittivity material. Sets of two electrode connectors (A and B) are brought into contact with each element to enable inter-element connections.

\subsection*{EM simulations}

Electromagnetic simulations were performed in CST (CST Studio Suite 2023, Dassault Systèmes, France) to model the $B_1^+$ field of a 3T MRI system. A 16-rung high-pass birdcage coil (diameter = 73 cm, length = 56 cm) was driven in circular polarization mode (i.e. applying a 22.5$^\circ$ phase shift between each successive rung). The coil was shielded by a copper cylindrical sheet (diameter = 79 cm, length = 100 cm).

A disk-shaped mineral oil phantom (diameter = 36 cm, height = 10 cm, $\epsilon_{r}$ = 2.1, $\sigma$ = 0.001 S/m)) was placed horizontally in the center of the coil. A 3$\times$3 array of dielectric blocks (width = 50 mm, height 30 mm) was placed on top of the phantom, aligned with the center. The default relative permittivity of the dielectric material was set to 165 unless stated otherwise, whereas the conductivity was set to $\sigma$ = 0.25 S/m in all simulations. For electrode connections, two copper rods (diameter = 2.5 mm, length = 37.5 mm) were placed in the top-left and bottom-right corner of each dielectric block at a 5 mm offset from the edge and extending down to 3 mm above the bottom of each block.

A time domain solver was set up to solve for the electromagnetic fields at 128 MHz in the simulation model. A non-uniform hexahedral mesh with minimal spacing of 3.5 mm around the phantom and coarser spacing outside the phantom was used for sufficient sampling of the gaps between the dielectric elements while minimizing the computation times. The simulated $B_1^+$ field was normalized to 1000 W of stimulated power and evaluated at 15 mm below the phantom surface.

Five different serial and parallel circuits between subsets of the dielectric array were studied to determine which interconnections produce a strong modulation of the $B_1^+$ field. Circuit connections were simulated by using perfectly conducting wires with a length of 6-11 cm (minimal connector-to-connector length, rounded to the nearest larger integer) and a radius of 0.2 mm. For parallel circuits, three versions were investigated: one connecting three adjacent elements (1A-2A-5A-1A and 1B-2B-5B-1B connections); four adjacent elements (1A-2A-5A-4A-1A and 1B-2B-5B-4B-1B); and four non-adjacent elements, with an additional gap (1A-3A-9A-7A-1A and 1B-3B-9B-7B-1B). For serial circuits, two versions were simulated: adjacent four-element connection (1A-2A, 2B-5A, 5B-4B, and 4A-1B) and adjacent two-element connection (1A-2A and 1B-2B). Longer wire lengths were also studied by simulating 4 cm, 8 cm and 12 cm extensions to the individual connections.

To quantify the modulation strength, $B_1^+$ differences were measured between the wired configurations and one reference configuration without any connections. For each configuration, a circular region-of-interest (ROI) was defined, centered at the point with maximum absolute modulation, with a radius of 11.7 mm to match the further phantom experiments. The configuration that induced the maximal change in $B_1^+$ was selected for use in further simulations and phantom experiments.

Subsequently, different subsets of array elements were connected using the optimized interconnections and compared to the uncoupled case in order to determine if the shimming effect can be spatially shifted. Finally, the relative permittivity of the dielectric was studied in order to evaluate the impact of dielectric properties on the device performance. In particular, $\epsilon_r$ was varied to resemble water ($\epsilon_r$ = 80), 25$\%$ volume barium titanate suspension in water (v/v) ($\epsilon_r$ = 165)\cite{teeuwisse2012, webb2011}, compressed metal titanates ($\epsilon_r$ = 500) \cite{Neves2018}, and ceramics ($\epsilon_r$ = 1000) \cite{koolstra2017,webb2022}. Finally, a low relative permittivity ($\epsilon_r$ = 10) and a few intermediate values ($\epsilon_r$ = 210, 255, 300) were evaluated.

\subsection*{Experimental prototype}

A prototype device was constructed using polylactic acid (PLA) on a commercially available 3D-printer (Sigma D25, BCN3D Technologies, Spain). A thin layer of epoxy resin was applied to the inner faces of the container to prevent water evaporation. Each cavity was filled with a 25$\%$ v/v suspension of barium titanate in distilled water. This ratio was chosen for its known dielectric properties at 3T ($\epsilon_r$ = 165), close to the saturation limit \cite{webb2011}.

After filling, the container was sealed with a 2 mm thick 3D-printed PLA lid that was mounted using a universal adhesive. The inner side of the lid was also coated with epoxy resin for waterproofing. A pair of header pins (PRPC040SACN-RC, Sullins Connector Solutions, United States) was inserted at two opposing corners of each dielectric element, at a distance of 5 mm to the lateral edges. A 27 mm copper wire was lowered into the dielectric to enable the electrical connections. The prototype is depicted in Fig.~\ref{fig1}C.

\subsection*{RF switching}

MRI-compatible RF switches were designed to modulate the RF current generated in an interconnected subset of the dielectric elements. MR-compatible PIN diodes (MA4P7470F-1072T, MACOM Technology Solutions) were used to switch interconnections based on the applied bias voltage. RF chokes (1 $\mu$H, 05CCM-1R0M-01, Fastron Group, Germany) and DC blocking capacitors (10 nF, VJ0402Y103KCAAC, Vishay, United States) were used to minimize RF coupling to the circuitry and direct the bias voltage to a single diode. The bias voltage was provided from the control room by a function generator (AFG31022, Tektronix, United States) connected to the RF switches in the scanner.

To study the optimized configuration, the RF switch was designed to interconnect four connecting points on the prototype. Each switch featured four PIN diodes, four header pins, eight DC chokes, and eight RF chokes (Fig.~\ref{fig1}B). The four PIN diode connections were grouped for simultaneous application of the bias voltage. The four header pins were connected to the container using 10 cm AWG 26 jumper wires. For connecting 2$\times$2 shim array subsets of adjacent elements in a parallel circuit (e.g. 1A-2A-5A-4A-1A and 1B-2B-5B-4B-1B for connecting elements 1, 2, 4, and 5), two switches were used.

\subsection*{$B_1^+$ mapping}

Phantom experiments were performed on a 3T MR system (Ingenia, Philips Healthcare, the Netherlands). $B_1^+$ maps were obtained using a Dual Refocusing Echo Acquisition Mode (DREAM) sequence (3x3 mm$^2$ in-plane resolution, 10 mm slice thickness, FOV = 450$\times$450 mm$^2$, TR/TE = 4.6/1.9 ms, Imaging/STEAM flip angle = 15/60 $^\circ$) \cite{nehrke2012}. RF shimming was evaluated in a disk-shaped phantom (diameter = 36 cm, height = 10 cm) filled with mineral oil (SpectraSyn 4, ExxonMobil, United States) and a ballistic gel torso (51$\times$46$\times$23 cm$^3$, T$_1$ $\approx$ 169~ms, T$_2$ $\approx$ 39.2~ms, Clear Ballistics, Unites States), which was used to mimic the human torso dimensions. $B_1^+$ maps were acquired approximately 15 mm below the surface of both phantoms.

First, the effect of the bias voltage was studied, acquiring $B_1^+$ maps in the cylindrical phantom for the bottom-left 2$\times$2 block connection, corresponding to dielectric elements 4, 5, 7, and 8. The bias voltage was varied between -5 V and 5 V, with 41 sampling steps, with a higher sampling density around the PIN diode activation between 0-1 V. One repetition per bias voltage was performed, and the $B_1^+$ modulation was evaluated in a small circular ROI with 11.7 mm radius (equivalent to 10 voxels in the reconstructed image).

Next, translation of the interconnection was studied by testing different circuit placements: connecting elements 1, 2, 4, and 5; elements 2, 3, 5, and 6; elements 4, 5, 7, and 8; elements 5, 6, 8, and 9. For this, a 5 V bias voltage was applied, comparing $B_1^+$ modulation to a reference case of 0 V. The maps were acquired with ten repetitions for each bias voltage in a cylindrical phantom and gel-torso, and the shimming effect was evaluated in each case over the small circular ROIs displaying the strongest modulation. The translatability was evaluated by estimating 2D correlation coefficient for the $B_1^+$ modulation area right below the connected 2$\times$2 array subset. Additionally, paired sample t-tests were performed for mean modulation within the ROIs, considering the Bonferroni correction-adjusted significant level of 0.01.

Finally, the $B_1^+$ modulation was studied as a function of depth in the gel-torso. $B_1^+$ maps were acquired at 5, 10, 15, 20, 25, 30, 35, 45, 55, 65, 75, 85, and 95 mm depths with a 5 V bias voltage applied to the PIN diodes. A reference map with 0 V bias voltage was also obtained for each slice. Three repetitions per slice were acquired for each bias voltage and depth. The maps were acquired using elements 2, 3, 5, and 6 connection as a representative configuration. For every slice, the $B_1^+$ modulation was evaluated for a small circular ROI.

\section*{Results}

\subsection*{EM simulations}

\begin{figure*}[ht]
    \centerline{\includegraphics[width=\textwidth]{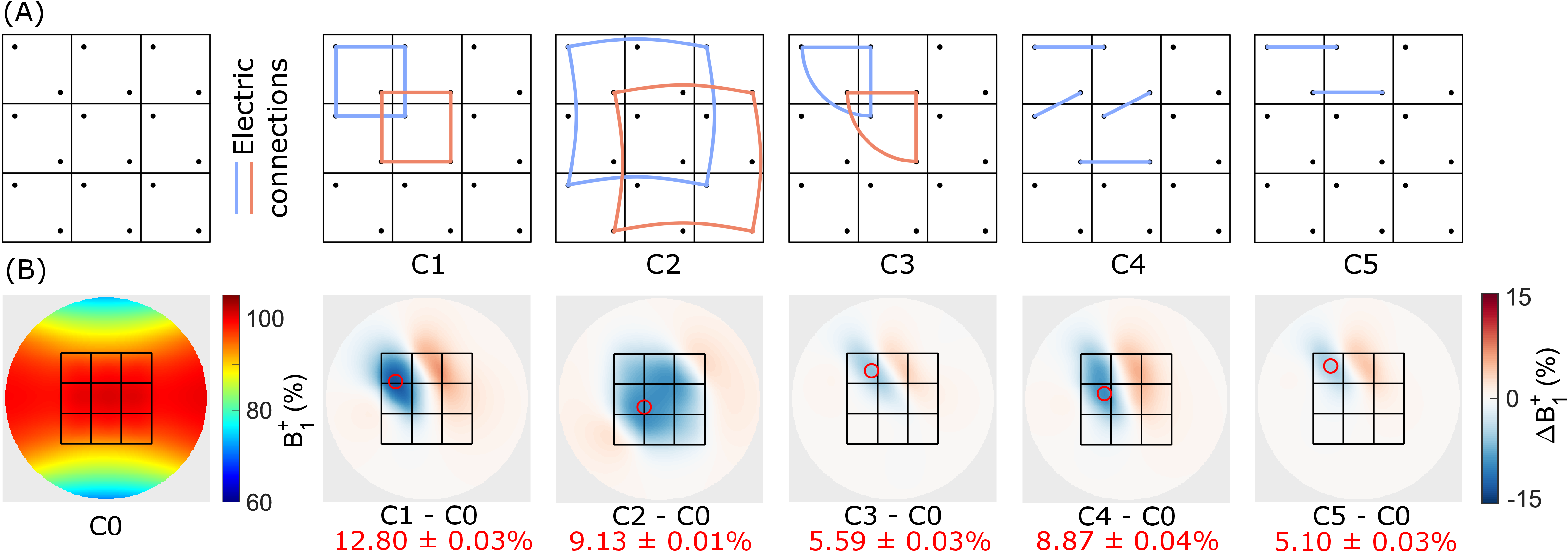}} 
    \caption{Simulated $B_1^+$ modulation for several interconnection configurations. (A) Schematic representation of a 3$\times$3 array of dielectric elements with different interconnection configurations. C0 represents the reference case without any interconnections. (B) $B_1^+$ difference maps evaluated at a depth of 15 mm inside the cylindrical phantom. The first subplot shows an absolute $B_1^+$ map for the reference configuration (C0, center value defined as 100\%). Subsequent subplots show the $B_1^+$ difference of each of the interconnected configurations (C1-C5) compared to the reference (C0). The outline of the dielectric array is shown in black, and the measurement regions of interest (ROIs) are indicated by a red circle.}
    \label{fig2}
\end{figure*}

EM simulations for several interconnection configurations show statistically significant $B_1^+$ field modulation (p\textless0.001, see Fig.~\ref{fig2}). A peak modulation of up to 12.8$\%$ is observed for four neighboring elements connected in a parallel circuit (C1). Alternative configurations with adjacent element connections (C3-C5) achieved a relatively weak modulation of up to 5.6$\%$, 8.9$\%$, and 5.1$\%$, respectively. Parallel configuration with non-adjacent connections of four blocks (C2) achieved 9.2$\%$ modulation, albeit much less localized. Notably, the modulation produced by all configurations studied was predominantly negative. Simulations with extended connection wire lengths showed a reduced $B_1^+$ field modulation, indicating that interconnecting wire length should be kept minimal (See the supplementary Fig. S1).

\begin{figure}[ht]
    \centerline{\includegraphics[width=0.8\textwidth]{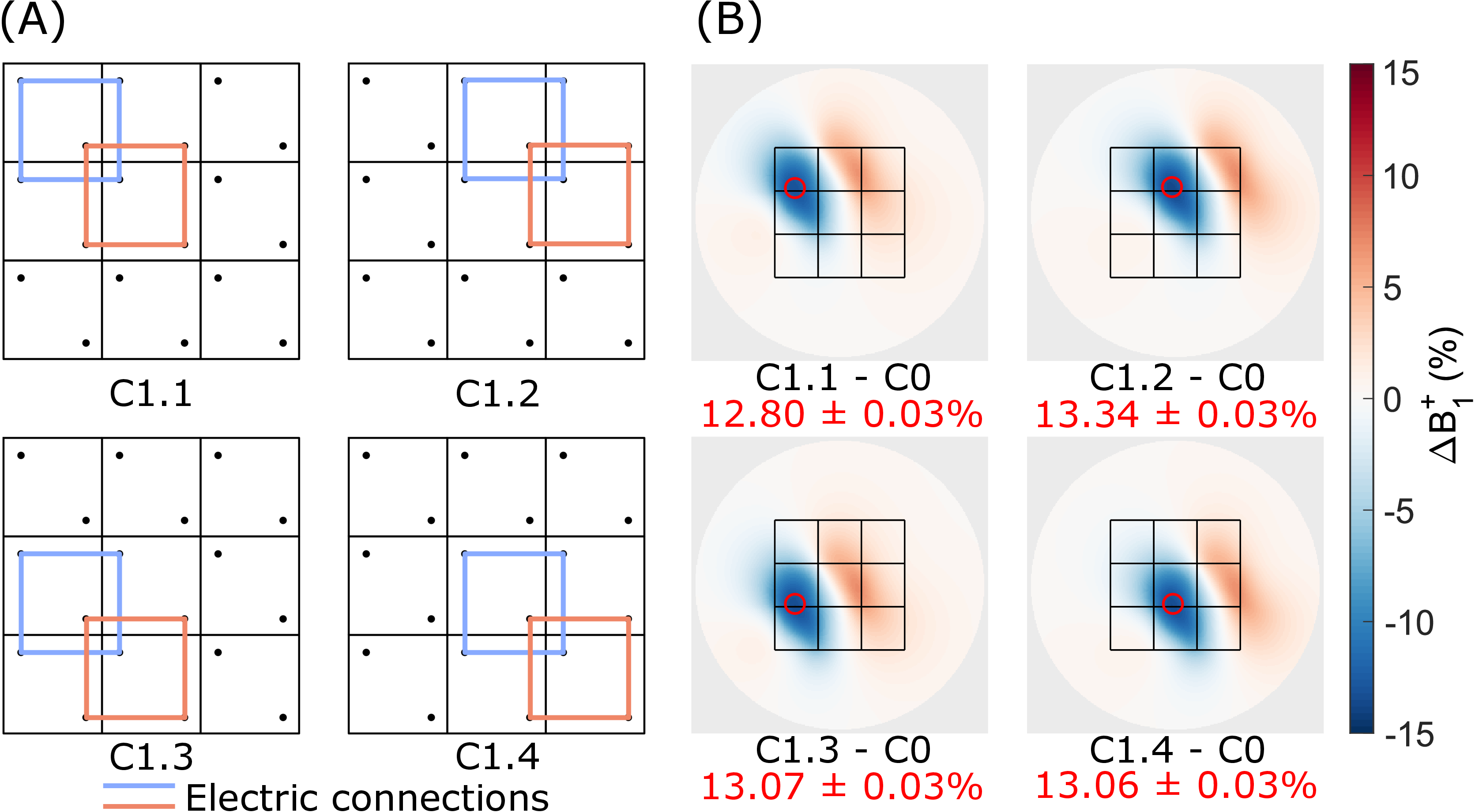}} 
    \caption{Simulated $B_1^+$ modulation produced by the C1 configuration translated within the array of dielectric blocks. (A) Schematic representation of C1 evaluated for four subsets of 2$\times$2 blocks in the array. (B) Normalized modulation of the $B_1^+$ field plotted as a difference between the coupled configuration and the uncoupled reference C0.}
    \label{fig3}
\end{figure}

The optimized circuit (C1) was shifted to multiple positions across the dielectric array, to study the spatial translation of the $B_1^+$ modulation as shown in Fig.~\ref{fig3}. Upon moving the circuit connection to a different 2$\times$2 subset of blocks (C1.1-C1.4), good translation of $B_1^+$ modulation pattern was observed. When comparing the $B_1^+$ modulation obtained in C1.1 directly below the corresponding elements (i.e. the area below elements 1, 2, 4, and 5) with that obtained in C1.2 - C1.4, the correlation coefficients are 0.9982, 0.9998, and 0.9988, respectively. The area of maximum $B_1^+$ field modulation was observed directly underneath the coupled blocks, reaching 12.8$\%$, 13.3$\%$, 13.1$\%$, and 13.1$\%$ modulation for C1.1-C1.4 configurations, respectively.

\begin{figure}[ht]
    \centerline{\includegraphics[width=0.8\textwidth]{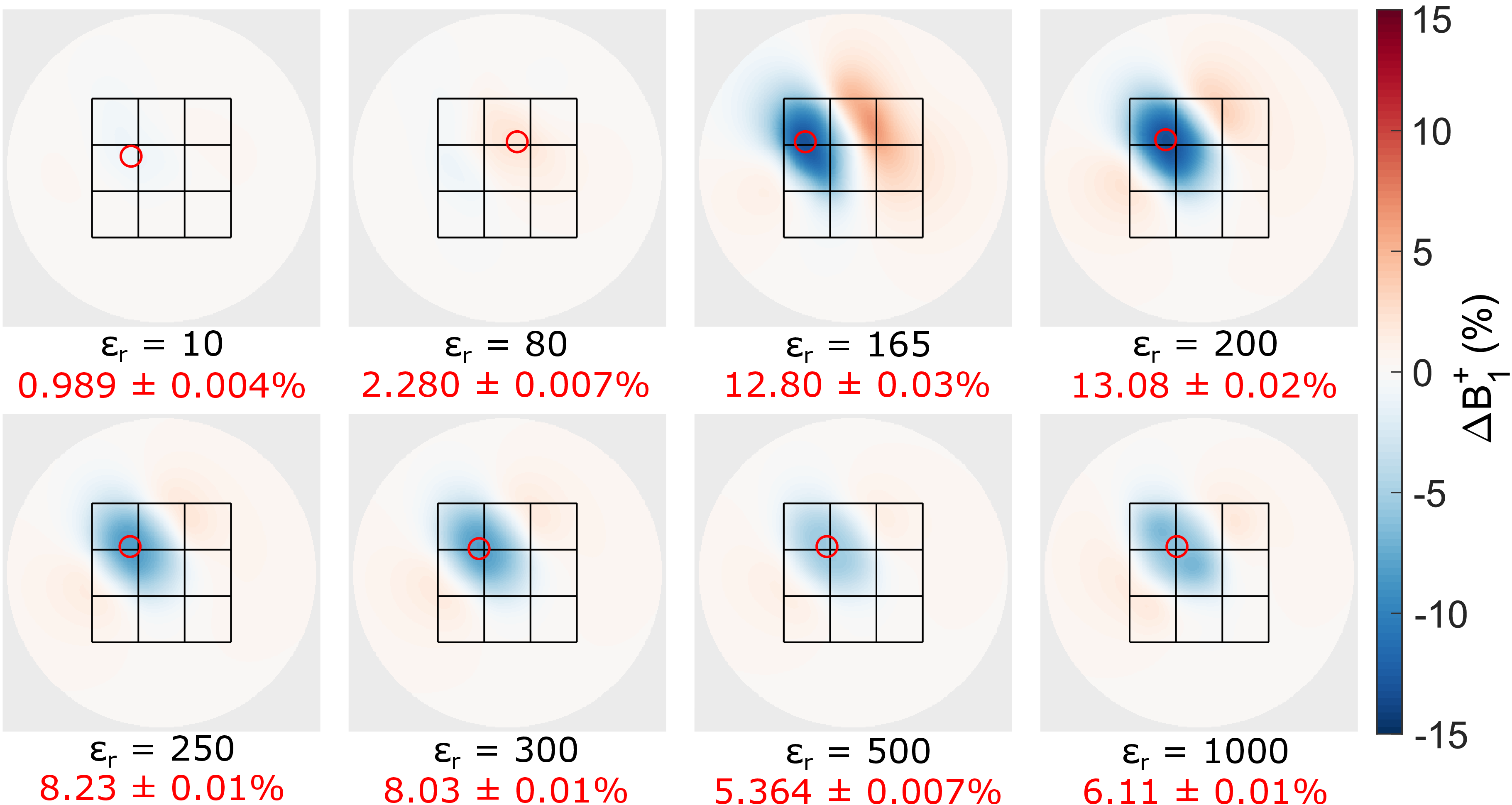}} 
    \caption{Simulated $B_1^+$ modulation showing the difference between configuration C1.1 and an uncoupled case C0 for various relative permittivity values of the dielectric elements.}
    \label{fig4}
\end{figure}

The $B_1^+$ modulation achieved by the shimming array showed a strong dependence on the relative permittivity of the dielectric material, as illustrated for the C1.1 configuration in Fig. \ref{fig4}. A low-permittivity ($\epsilon_r = 10$) and water model ($\epsilon_r = 80$) result in a weak $B_1^+$ modulation of 0.99$\%$ and 2.28$\%$, respectively. Maximum modulations of 13.1$\%$ and 12.8$\%$ are achieved by $\epsilon_r = 200$ and $\epsilon_r = 165$. Increasing the relative permittivity value above 250 resulted in a decreased modulation.

\subsection*{Phantom results}

\begin{figure}[ht]
    \centerline{\includegraphics[width=0.7\textwidth]{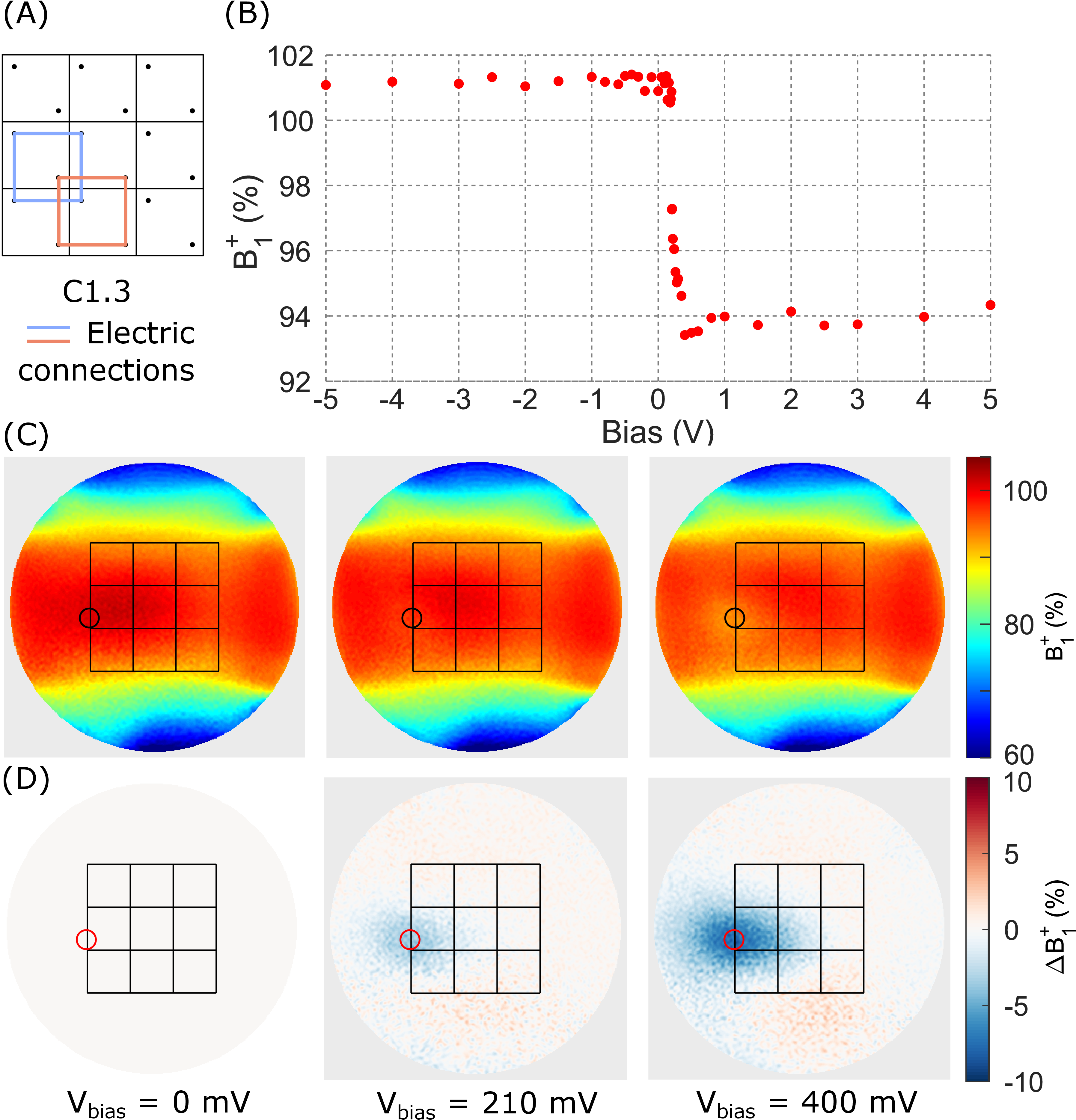}} 
    \caption{Adaptive $B_1^+$ modulation of the dielectric array measured in a disk-shaped phantom using C1.3 configuration. (A) $B_1^+$ (percentage of nominal value) in a circular ROI plotted as a function of applied bias voltage $V_{bias}$, showing increased $B_1^+$ modulation at high $V_{bias}$ as the resistance across the PIN diodes decreases. (C) Normalized $B_1^+$ maps for three bias voltages. The circular ROI used to extract the data in (A) is indicated in black. (C) Relative $B_1^+$ maps showing the difference between the $B_1^+$ maps of (B) and a reference case of 0 V bias voltage, with the ROI indicated by a red circle.}
    \label{fig5}
\end{figure}

The $B_1^+$ modulation was evaluated as a function of the bias voltage $V_{bias}$ in the disk-shaped phantom by gradually increasing the bias voltage $V_{bias}$ across the PIN diodes (see Fig. \ref{fig5}). A steep onset was observed around $V_{bias} = $200-400~mV, reaching a plateau at around 800 mV with mean $B_1^+$ of 93.94$\pm$0.04$\%$. Negligible variation is observed for negative bias voltages (mean normalized $B_1^+$ of 101.18$\pm$0.03$\%$). Accordingly, the configuration C1.3 achieves up to 7.24$\pm$0.03$\%$ $B_1^+$ modulation.

\begin{figure*}[ht]
    \centerline{\includegraphics[width=0.8\textwidth]{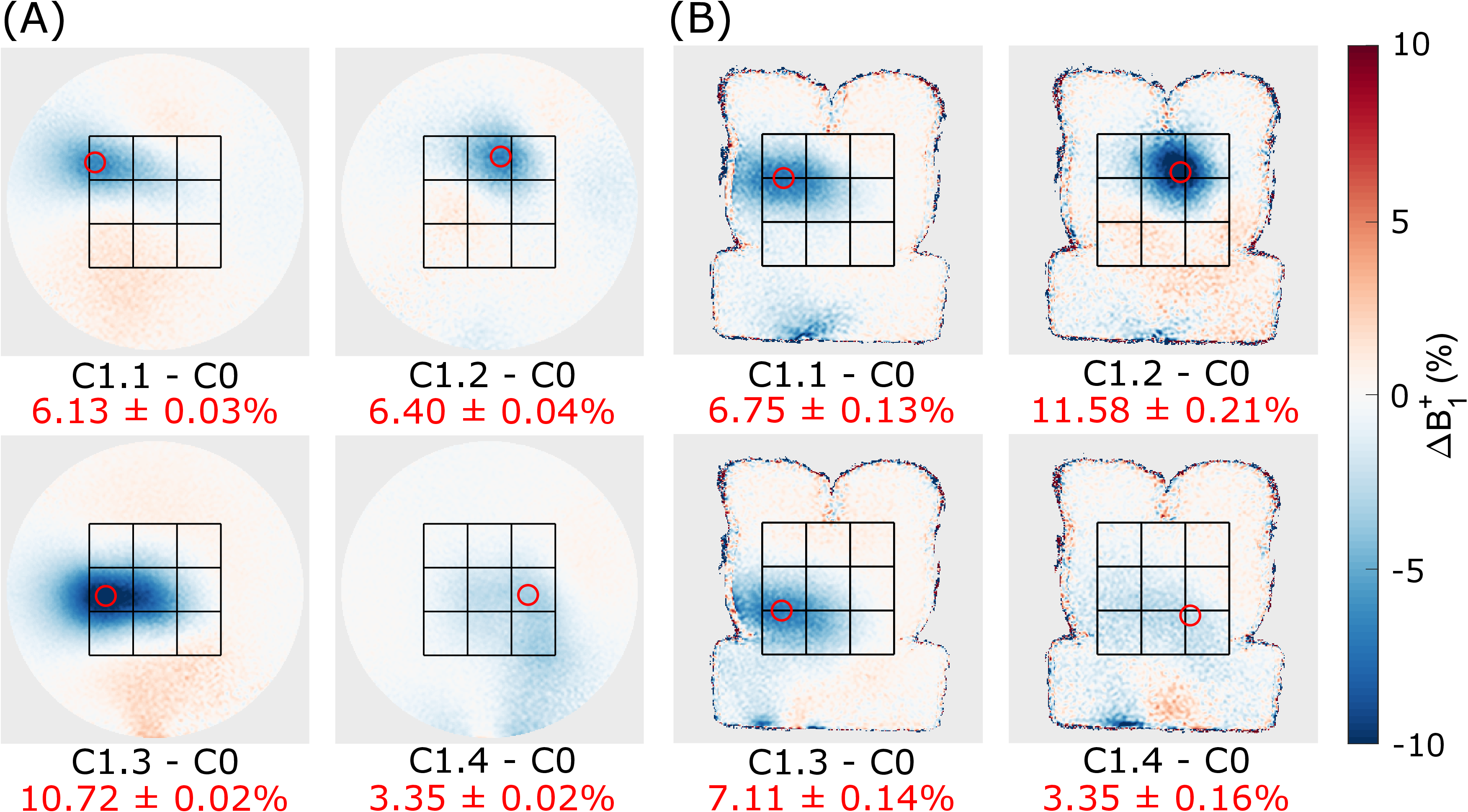}} 
    \caption{Modulation of the $B_1^+$ field in the two phantoms produced by the C1 configuration translated within the dielectric shimming array. The maps are shown as a difference between the 5 V bias and a 0 V reference state (C0). (A) Modulation patterns in the disk-shaped phantom. (B) Modulation patterns in the torso phantom. The coupling configuration and modulation in the ROIs (red circle) is stated below each subplot.}
    \label{fig6}
\end{figure*}

Similar to simulations, the $B_1^+$ modulation in both the disk-shaped and gel-torso phantoms showed consistent spatial translation, when different 2$\times$2 subsets of blocks were connected in configuration C1. Larger differences in the modulation patterns between the different subsets were observed in the experiments compared to simulations. Accordingly, a lower correlation coefficients between the modulation patterns for C1.2-C1.4 and C1.1 as reference was observed in the cylindrical phantom (C1.2 vs C1.1: 0.896, C1.3 vs C1.1: 0.961, and C1.4 vs C1.1: 0.762). In C1.1-C1.4, maximum $B_1^+$ modulation of 6.13$\pm$0.03$\%$, 6.40$\pm$0.04$\%$, 10.72$\pm$0.02$\%$, and 3.35$\pm$0.02$\%$ was observed for the disk-shaped phantom, respectively (see Fig.~\ref{fig6}A, p\textless0.001). In the gel-torso phantom, comparable $B_1^+$ modulation was achieved reaching 6.75$\pm$0.13$\%$, 11.58$\pm$0.22$\%$, 7.11$\pm$0.14$\%$, and 3.35$\pm$0.16$\%$ for C1.1-C1.4 (see Fig.~\ref{fig6}B, p\textless0.001). Correlation coefficients upon translation of C1.2-C1.4 with respect to C1.1 were 0.887, 0.986, and 0.929, respectively.

\begin{figure*}[ht]\centerline{\includegraphics[width=0.8\textwidth]{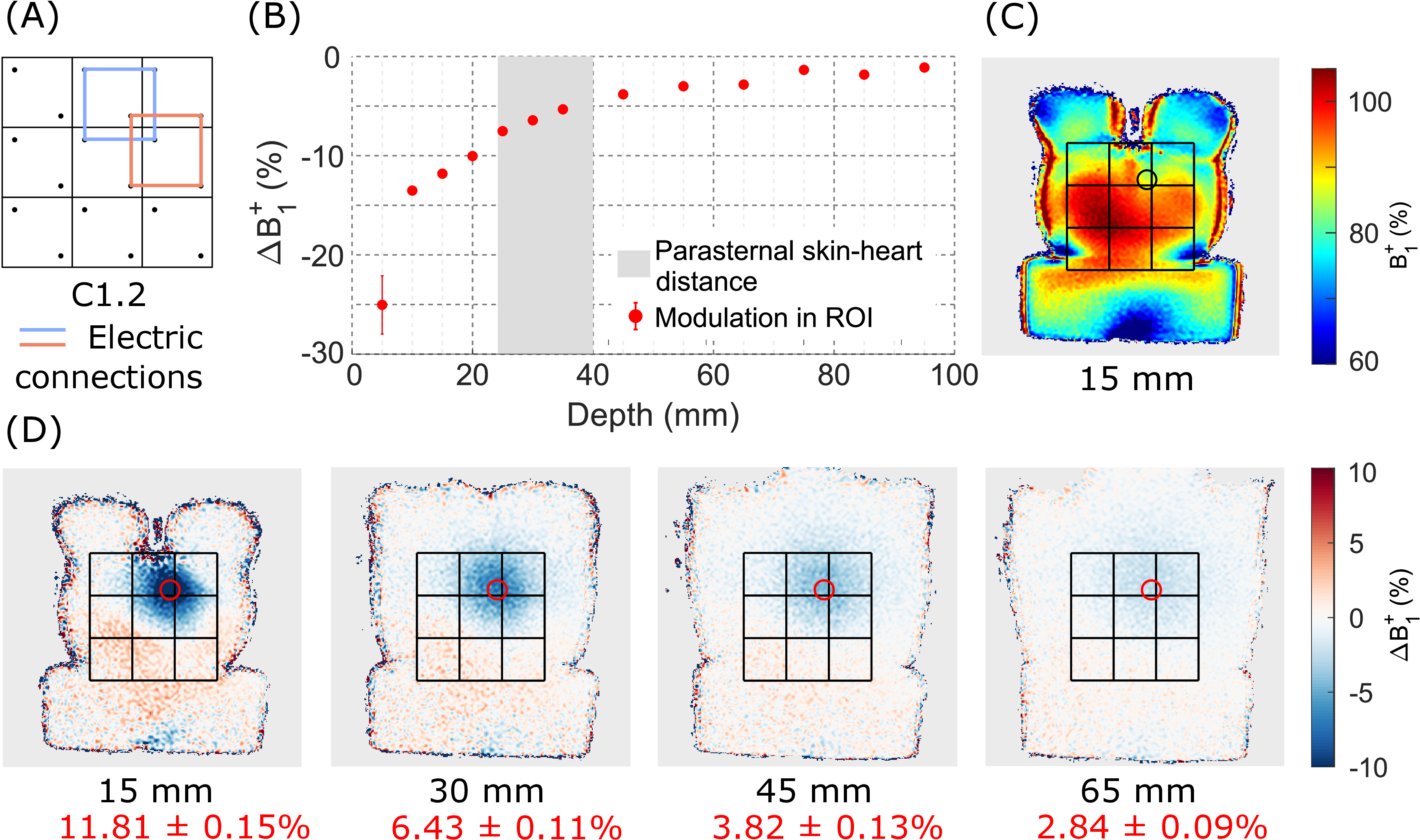}} 
    \caption{$B_1^+$ shimming effect in the gel-torso phantom for slices at various depths under the dielectric array. (A) The top-right 2$\times$2 blocks were interconnected (C1.2) using a 5 V forward bias voltage. (B) $B_1^+$ modulation as a function of imaging depth, measured in the ROI indicated by a black or red circle in (C-D). The shaded gray area shows the parasternal skin-to-heart distance within one standard deviation from the mean distance \cite{rahko2008}. (C) Absolute $B_1^+$ map acquired 15 mm below the phantom surface. (D) $B_1^+$ modulation for slices at exemplary depths.}
    \label{fig7}
\end{figure*}

The $B_1^+$ variation with depth within the gel-torso phantom showed a gradual decay of the $B_1^+$ modulation when moving further away from the dielectric array (see Fig.~\ref{fig7}). The $B_1^+$ modulation reaches up to 25.0$\pm$2.9$\%$ near the phantom surface, and is still measurable at 95 mm below the surface (1.12$\pm$0.05$\%$). At 25-40 mm depth, which is typical for cardiac anatomy, 4-8$\%$ modulation is achieved.

\section*{Discussion}

In this study, a proof-of-principle device for adaptive $B_1^+$ field shimming in MRI was built using an array of interconnected dielectric elements. EM simulations indicated that shifting a set of parallel connections across the shimming array enables targeted spatial modulation of the $B_1^+$ field. Accordingly, phantom experiments using PIN diode-based RF switches between elements filled with dielectric slurry show significant and controllable spatial modulation of the $B_1^+$ field. This demonstrates a promising proof-of-principle for realizing adjustable $B_1^+$ shimming effects using dielectric materials, as a cost-effective solution for $B_1^+$ shimming in high field MRI.

From the investigated set of different interconnection configurations in EM simulations, four adjacent blocks connected in parallel were chosen for the phantom experiments due to the achieved focal field modulation. Notably, the configuration produced predominantly negative $B_1^+$ field modulation, i.e. reducing the total $B_1^+$ field amplitude. This could prove valuable in reducing RF-induced heating around implants \cite{jacobs2024ismrm}, or to homogenize the transmit field by locally suppressing hyper-enhanced $B_1^+$ areas. However, in order to correct areas affected by signal dropout, a positive $B_1^+$ field modulation would be more beneficial in terms of RF power efficiency. Simulation results indicate that various device properties, such as the exact permittivity of the dielectric material, or the length of the coupling wires can substantially change the modulation pattern, potentially shifting from negative to positive modulation. In addition to the modulation polarity, the absolute magnitude of the induced $B_1^+$ field modulation also warrants further investigation. In the current design, up to 11$\%$ absolute modulation was observed at a depth of 15 mm below the device. Considering the typical levels of $B_1^+$ inhomogeneity ranging from 30-50 $\%$ for body imaging at 3T \cite{sung2008}, further improvements in the shimming magnitude are needed to achieve sufficient RF field control in clinical applications. Future device optimization will therefore include tuning of the dielectric properties, as well as the geometry of the dielectric elements and interconnections, in order to fully exploit the potential for $B_1^+$ field correction.

PIN diode switches were used to enable control of the $B_1^+$ modulation magnitude in this work, operated either in reverse or forward bias. Based on the bias voltage analysis, where a smooth modulation magnitude function was observed, one might consider driving the diodes close to the activation voltage, to provide intermediate levels of $B_1^+$ field modulation. While PIN diodes proved to be effective, metal–oxide–semiconductor field-effect transistors (MOSFETs) or microelectromechanical systems (MEMS) switches can be further explored as MR-compatible alternatives to enable RF switching functionalities. These alternatives may benefit from decreased power consumption, potentially enabling completely wireless implementations of the device. Furthermore, MOSFET or MEMS switches can also reduce ohmic losses in the device and improve efficiency, as it requires a fewer total number of components.

In the current prototype, the $B_1^+$ field modulation was optimized for modulation magnitude rather than total $B_1^+$ homogeneity. Improving this requires further tailoring of the element combinations. In an experimental setting, one might start by mapping the initial $B_1^+$ field, followed by semi-automated activation of dielectric elements close to the region of interest. Iterative application of $B_1^+$ mapping and adjustment of the switching modes can allow for robust and adaptive RF field homogenization in a subject-specific manner. The duration of such a calibration procedure would be largely determined by the scan time required for $B_1^+$ mapping, which can be achieved within seconds \cite{nehrke2012,siuryte2024preparation}, potentially allowing for automated calibration within a clinically acceptable time-frame of approximately 30 seconds.

Various design improvements are warranted to bring the prototype closer to clinical use. For body and cardiac MRI, larger arrays with an increased number of elements should be considered to provide sufficient coverage and increased degrees of freedom for tailoring the spatial modulation pattern. Furthermore, a flexible body-conforming array can be constructed to improve both patient comfort and effect size due to a decreased separation between the dielectric and the skin. Such a device could also be integrated with receive coil geometries, such as chest arrays or head coils. This would be particularly useful for brain imaging at ultra-high field strengths such as 7T, where the $B_1^+$ field exhibits even stronger inhomogeneity patterns compared to 3T. To ensure that the device fits the limited space inside the close-fitting head coil, thinner dielectric elements with increased permittivity can be explored.

The current work has several limitations. Firstly, only a small number of potential electrical connections was studied. The current results warrant further investigation to explore additional interconnections with different layouts, the individual dielectric element properties, and the achieved $B_1^+$ modulation pattern. Furthermore, in EM simulations, the $B_1^+$ field modulation was only studied in the oil phantom, which can be extended to anatomical body models for further optimization of the effects. Various human body models could be used to establish robust and reproducible shimming across a range of subjects. Finally, the current prototype has only been evaluated in phantom imaging. Further in-vivo studies will be needed to validate the effectiveness and quantify the achieved $B_1^+$ modulation, while ascertaining safety in terms of RF-induced heating due to the device and adjacent tissue \cite{brink2023}.

\section*{Conclusion}

In this work, a cost-effective method for adaptive modulation of the $B_1^+$ field in high field MRI using reconfigurable dielectric materials is proposed. The proof-of-principle demonstration shows reproducible and localized modulation of the $B_1^+$ field in simulations and phantom experiments.
A $B_1^+$ field modulation magnitude of up to 11\% was observed in phantom measurements 15 mm below the surface. Further optimization of the material properties and the design parameters is warranted to increase the effectiveness further. This provides a promising alternative towards adaptive and cost-effective $B_1^+$ shimming without the need for multiple transmit coils and associated hardware and software requirements.

\bibliography{main_R0}

\begin{thebibliography}{10}
\urlstyle{rm}
\expandafter\ifx\csname url\endcsname\relax
  \def\url#1{\texttt{#1}}\fi
\expandafter\ifx\csname urlprefix\endcsname\relax\def\urlprefix{URL }\fi
\expandafter\ifx\csname doiprefix\endcsname\relax\def\doiprefix{DOI: }\fi
\providecommand{\bibinfo}[2]{#2}
\providecommand{\eprint}[2][]{\url{#2}}

\bibitem{bernstein2006}
\bibinfo{author}{Bernstein, M.~A.}, \bibinfo{author}{Huston~III, J.} \& \bibinfo{author}{Ward, H.~A.}
\newblock \bibinfo{journal}{\bibinfo{title}{Imaging artifacts at 3.0 t}}.
\newblock {\emph{\JournalTitle{Journal of Magnetic Resonance Imaging: An Official Journal of the International Society for Magnetic Resonance in Medicine}}} \textbf{\bibinfo{volume}{24}}, \bibinfo{pages}{735--746} (\bibinfo{year}{2006}).

\bibitem{kataoka2007}
\bibinfo{author}{Kataoka, M.} \emph{et~al.}
\newblock \bibinfo{journal}{\bibinfo{title}{Mr imaging of the female pelvis at 3 tesla: evaluation of image homogeneity using different dielectric pads}}.
\newblock {\emph{\JournalTitle{Journal of Magnetic Resonance Imaging: An Official Journal of the International Society for Magnetic Resonance in Medicine}}} \textbf{\bibinfo{volume}{26}}, \bibinfo{pages}{1572--1577} (\bibinfo{year}{2007}).

\bibitem{soher2007}
\bibinfo{author}{Soher, B.~J.}, \bibinfo{author}{Dale, B.~M.} \& \bibinfo{author}{Merkle, E.~M.}
\newblock \bibinfo{journal}{\bibinfo{title}{A review of mr physics: 3t versus 1.5 t}}.
\newblock {\emph{\JournalTitle{Magnetic resonance imaging clinics of North America}}} \textbf{\bibinfo{volume}{15}}, \bibinfo{pages}{277--290} (\bibinfo{year}{2007}).

\bibitem{sung2008}
\bibinfo{author}{Sung, K.} \& \bibinfo{author}{Nayak, K.~S.}
\newblock \bibinfo{journal}{\bibinfo{title}{Measurement and characterization of rf nonuniformity over the heart at 3t using body coil transmission}}.
\newblock {\emph{\JournalTitle{Journal of Magnetic Resonance Imaging: An Official Journal of the International Society for Magnetic Resonance in Medicine}}} \textbf{\bibinfo{volume}{27}}, \bibinfo{pages}{643--648} (\bibinfo{year}{2008}).

\bibitem{gutberlet2004}
\bibinfo{author}{Gutberlet, M.} \emph{et~al.}
\newblock \bibinfo{title}{Comparison of different cardiac mri sequences at 1.5 t/3.0 t with respect to signal-to-noise and contrast-to-noise ratios-initial experience}.
\newblock In \emph{\bibinfo{booktitle}{R{\"o}Fo-Fortschritte auf dem Gebiet der R{\"o}ntgenstrahlen und der bildgebenden Verfahren}}, vol. \bibinfo{volume}{176}, \bibinfo{pages}{801--808} (\bibinfo{organization}{{\copyright} Georg Thieme Verlag KG Stuttgart{\textperiodcentered} New York}, \bibinfo{year}{2004}).

\bibitem{deniz2019}
\bibinfo{author}{Deniz, C.~M.}
\newblock \bibinfo{journal}{\bibinfo{title}{Parallel transmission for ultrahigh field mri}}.
\newblock {\emph{\JournalTitle{Topics in Magnetic Resonance Imaging}}} \textbf{\bibinfo{volume}{28}}, \bibinfo{pages}{159--171} (\bibinfo{year}{2019}).

\bibitem{brink2015a}
\bibinfo{author}{Brink, W.~M.}, \bibinfo{author}{Gulani, V.} \& \bibinfo{author}{Webb, A.~G.}
\newblock \bibinfo{journal}{\bibinfo{title}{Clinical applications of dual-channel transmit mri: A review}}.
\newblock {\emph{\JournalTitle{Journal of Magnetic Resonance Imaging}}} \textbf{\bibinfo{volume}{42}}, \bibinfo{pages}{855--869} (\bibinfo{year}{2015}).

\bibitem{wald2009}
\bibinfo{author}{Wald, L.~L.} \& \bibinfo{author}{Adalsteinsson, E.}
\newblock \bibinfo{journal}{\bibinfo{title}{Parallel transmit technology for high field mri}}.
\newblock {\emph{\JournalTitle{Magnetom Flash}}} \textbf{\bibinfo{volume}{40}}, \bibinfo{pages}{2009} (\bibinfo{year}{2009}).

\bibitem{williams2023}
\bibinfo{author}{Williams, S.~N.}, \bibinfo{author}{McElhinney, P.} \& \bibinfo{author}{Gunamony, S.}
\newblock \bibinfo{journal}{\bibinfo{title}{Ultra-high field mri: parallel-transmit arrays and rf pulse design}}.
\newblock {\emph{\JournalTitle{Physics in Medicine \& Biology}}} \textbf{\bibinfo{volume}{68}}, \bibinfo{pages}{02TR02} (\bibinfo{year}{2023}).

\bibitem{vorobyev2020}
\bibinfo{author}{Vorobyev, V.} \emph{et~al.}
\newblock \bibinfo{journal}{\bibinfo{title}{An artificial dielectric slab for ultra high-field mri: Proof of concept}}.
\newblock {\emph{\JournalTitle{Journal of Magnetic Resonance}}} \textbf{\bibinfo{volume}{320}}, \bibinfo{pages}{106835} (\bibinfo{year}{2020}).

\bibitem{teeuwisse2012}
\bibinfo{author}{Teeuwisse, W.}, \bibinfo{author}{Brink, W.}, \bibinfo{author}{Haines, K.} \& \bibinfo{author}{Webb, A.}
\newblock \bibinfo{journal}{\bibinfo{title}{Simulations of high permittivity materials for 7 t neuroimaging and evaluation of a new barium titanate-based dielectric}}.
\newblock {\emph{\JournalTitle{Magnetic resonance in medicine}}} \textbf{\bibinfo{volume}{67}}, \bibinfo{pages}{912--918} (\bibinfo{year}{2012}).

\bibitem{webb2011}
\bibinfo{author}{Webb, A.~G.}
\newblock \bibinfo{journal}{\bibinfo{title}{Dielectric materials in magnetic resonance}}.
\newblock {\emph{\JournalTitle{Concepts in magnetic resonance part A}}} \textbf{\bibinfo{volume}{38}}, \bibinfo{pages}{148--184} (\bibinfo{year}{2011}).

\bibitem{koolstra2017}
\bibinfo{author}{Koolstra, K.}, \bibinfo{author}{B{\"o}rnert, P.}, \bibinfo{author}{Brink, W.} \& \bibinfo{author}{Webb, A.}
\newblock \bibinfo{journal}{\bibinfo{title}{Improved image quality and reduced power deposition in the spine at 3 t using extremely high permittivity materials}}.
\newblock {\emph{\JournalTitle{Magnetic resonance in medicine}}} \textbf{\bibinfo{volume}{79}}, \bibinfo{pages}{1192--1199} (\bibinfo{year}{2018}).

\bibitem{webb2022}
\bibinfo{author}{Webb, A.}, \bibinfo{author}{Shchelokova, A.}, \bibinfo{author}{Slobozhanyuk, A.}, \bibinfo{author}{Zivkovic, I.} \& \bibinfo{author}{Schmidt, R.}
\newblock \bibinfo{journal}{\bibinfo{title}{Novel materials in magnetic resonance imaging: high permittivity ceramics, metamaterials, metasurfaces and artificial dielectrics}}.
\newblock {\emph{\JournalTitle{Magnetic Resonance Materials in Physics, Biology and Medicine}}} \textbf{\bibinfo{volume}{35}}, \bibinfo{pages}{875--894} (\bibinfo{year}{2022}).

\bibitem{brink2015}
\bibinfo{author}{Brink, W.~M.}, \bibinfo{author}{Remis, R.~F.} \& \bibinfo{author}{Webb, A.~G.}
\newblock \bibinfo{journal}{\bibinfo{title}{A theoretical approach based on electromagnetic scattering for analysing dielectric shimming in high-field mri}}.
\newblock {\emph{\JournalTitle{Magnetic resonance in medicine}}} \textbf{\bibinfo{volume}{75}}, \bibinfo{pages}{2185--2194} (\bibinfo{year}{2016}).

\bibitem{brink2014}
\bibinfo{author}{Brink, W.~M.} \& \bibinfo{author}{Webb, A.~G.}
\newblock \bibinfo{journal}{\bibinfo{title}{High permittivity pads reduce specific absorption rate, improve b1 homogeneity, and increase contrast-to-noise ratio for functional cardiac mri at 3 t}}.
\newblock {\emph{\JournalTitle{Magnetic resonance in medicine}}} \textbf{\bibinfo{volume}{71}}, \bibinfo{pages}{1632--1640} (\bibinfo{year}{2014}).

\bibitem{heer2016}
\bibinfo{author}{de~Heer, P.}, \bibinfo{author}{Bizino, M.~B.}, \bibinfo{author}{Versluis, M.~J.}, \bibinfo{author}{Webb, A.~G.} \& \bibinfo{author}{Lamb, H.~J.}
\newblock \bibinfo{journal}{\bibinfo{title}{Improved cardiac proton magnetic resonance spectroscopy at 3 t using high permittivity pads}}.
\newblock {\emph{\JournalTitle{Investigative radiology}}} \textbf{\bibinfo{volume}{51}}, \bibinfo{pages}{134--138} (\bibinfo{year}{2016}).

\bibitem{Neves2018}
\bibinfo{author}{Neves, A.~L.} \emph{et~al.}
\newblock \bibinfo{journal}{\bibinfo{title}{Compressed perovskite aqueous mixtures near their phase transitions show very high permittivities: New prospects for high-field mri dielectric shimming}}.
\newblock {\emph{\JournalTitle{Magnetic resonance in medicine}}}  (\bibinfo{year}{2018}).

\bibitem{nehrke2012}
\bibinfo{author}{Nehrke, K.} \& \bibinfo{author}{B{\"o}rnert, P.}
\newblock \bibinfo{journal}{\bibinfo{title}{Dream—a novel approach for robust, ultrafast, multislice b1 mapping}}.
\newblock {\emph{\JournalTitle{Magnetic resonance in medicine}}} \textbf{\bibinfo{volume}{68}}, \bibinfo{pages}{1517--1526} (\bibinfo{year}{2012}).

\bibitem{jacobs2024ismrm}
\bibinfo{author}{Jacobs, P.~S.} \emph{et~al.}
\newblock \bibinfo{title}{Reduction of radiofrequency induced heating around passive implants via flexible metasurface shielding at 7t}.
\newblock In \emph{\bibinfo{booktitle}{Proceedings of the International Society for Magnetic Resonance in Medicine}}, vol. \bibinfo{volume}{1215} (\bibinfo{year}{2024}).

\bibitem{siuryte2024preparation}
\bibinfo{author}{{\v{S}}iuryt{\.e}, P.} \emph{et~al.}
\newblock \bibinfo{journal}{\bibinfo{title}{Preparation-based b1+ mapping in the heart using bloch-siegert shifts}}.
\newblock {\emph{\JournalTitle{Magnetic resonance in medicine}}}  (\bibinfo{year}{2024}).

\bibitem{brink2023}
\bibinfo{author}{Brink, W.~M.}, \bibinfo{author}{Remis, R.~F.} \& \bibinfo{author}{Webb, A.~G.}
\newblock \bibinfo{journal}{\bibinfo{title}{Radiofrequency safety of high permittivity pads in mri—impact of insulation material}}.
\newblock {\emph{\JournalTitle{Magnetic resonance in medicine}}} \textbf{\bibinfo{volume}{89}}, \bibinfo{pages}{2109--2116} (\bibinfo{year}{2023}).

\bibitem{rahko2008}
\bibinfo{author}{Rahko, P.~S.}
\newblock \bibinfo{journal}{\bibinfo{title}{Evaluation of the skin-to-heart distance in the standing adult by two-dimensional echocardiography}}.
\newblock {\emph{\JournalTitle{Journal of the American Society of Echocardiography}}} \textbf{\bibinfo{volume}{21}}, \bibinfo{pages}{761--764} (\bibinfo{year}{2008}).

\end{thebibliography}
e

\section*{Acknowledgements}

The authors would like to thank Dr.ir. Rob F. Remis for the initial project discussions.

\section*{Author contributions statement}

S.W. conceived the device working principle and research aims. 
S.W., O.C.A. and W.B. conceived the initial experimental design. 
S.W., P.S., J.L. and R.V. designed the experiments. 
J.L. and R.V. performed the electromagnetic field simulations.
P.S., J.L. and R.V. acquired the experimental MRI data.
R.V. conducted the data analysis.
S.W. contributed to data analysis and discussion. 
Manuscript figures were prepared by J.L. and R.V. with input from P.S. and S.W. 
The manuscript draft was was written by J.L., R.V. and P.S., and reviewed and edited by all authors.

\section*{Funding}

This project was co-funded by the Nederlandse Organisatie voor Wetenschappelijk Onderzoek (NWO) Start-up Grant, STU.019.024; the 4TU Precision Medicine program supported by High Tech for a Sustainable Future; and the European Union ERC StG, VascularID, 101078711.

\section*{Additional information}

\textbf{Accession codes} All experimental data can be found at \newline \url{https://gitlab.tudelft.nl/mars-lab/active-dielectric-shimming}; \textbf{Competing interests} O.C.A., S.W. and P.S. are inventors on a patent application related to this manuscript (OCT-21-092), submitted by Delft University of Technology.

\end{document}


\thispagestyle{empty}%
\vskip-36pt%
{\raggedright\sffamily\bfseries\fontsize{20}{25}\selectfont Adaptive Radiofrequency Shimming in MRI using Reconfigurable Dielectric Materials: \newline Supplementary Materials\par}%
\vskip10pt
{\raggedright\sffamily\fontsize{12}{16}\selectfont Paulina \v{S}iuryt\.{e}, Robert van de Velde, Jasper van Leeuwen, \"{O}mer Can Akg\"{u}n, Wyger Brink and Sebastian~Weing\"{a}rtner\par}
\vskip18pt%

    \begin{figure*}[ht]
        \centerline{\includegraphics[width=0.95\textwidth]{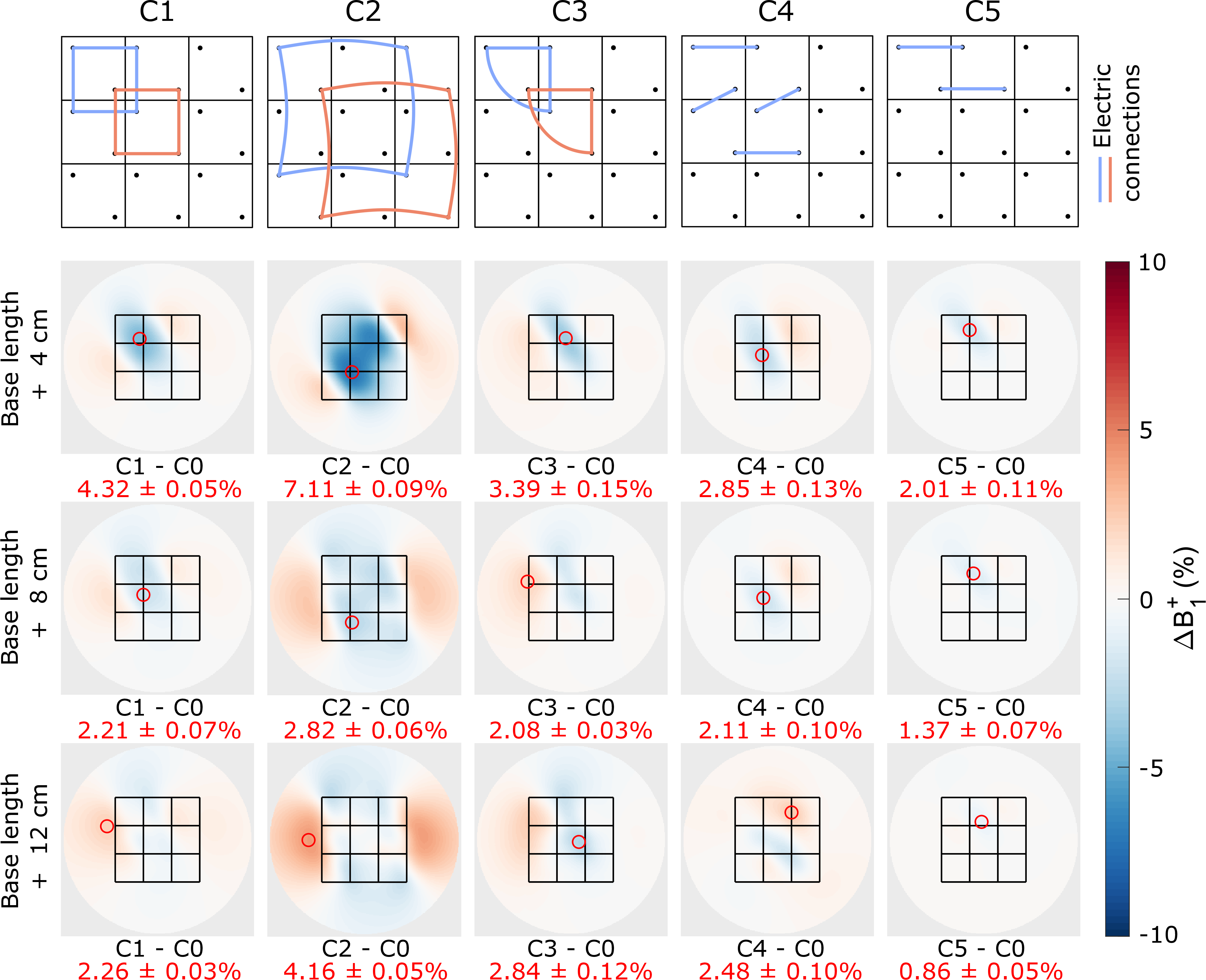}} 
        \caption{Simulated $B_1^+$ modulation for several studied coupling configurations. (A) Schematic representation of a 3$\times$3 array of dielectric pockets with different coupling configurations. C0 is uncoupled and is provided as a reference case. (B) $B_1^+$ difference maps evaluated at a depth of 15 mm inside the cylindrical phantom and normalized to 1000 W of stimulated power. Each subplot shows the difference in the normalized $B_1^+$ field of the coupled case (C1-C5) and the baseline C0. The outline of the cask is shown in black, and the considered ROIs are indicated by a red circle. The first subplot shows an absolute $B_1^+$ map for the uncoupled case C0.}
        \label{fig2}
    \end{figure*}